\documentclass{aa}

\usepackage{graphicx}
\usepackage{txfonts}
\usepackage{color}
\usepackage{url}
\usepackage[flushleft]{threeparttable}
\usepackage[colorlinks=true, allcolors=blue]{hyperref}
\usepackage{orcidlink}
\usepackage{adjustbox}
\usepackage{multicol}
\usepackage{multirow}
\usepackage{xspace}
\newcommand {\gx}{{GX~340$+$0}\xspace}
\newcommand {\ixpe}{{IXPE}\xspace}
\newcommand {\nustar}{\textit{NuSTAR}\xspace}

\newcommand {\nicer}{{NICER}\xspace} 
\newcommand {\ep}{{Einstein~Probe}\xspace} 
\newcommand{\fluxcgs}{erg\,
s$^{-1}$\,cm$^{-2}$}
\newcommand{\lum}{erg\,s$^{-1}$}

\begin{document}

\title{X-ray spectropolarimetric characterisation of the Z source GX~340$+$0 in the normal branch}

\titlerunning{X-ray spectropolarimetric characterisation of the Z source GX~340$+$0 in the normal branch}

\author{
Fabio La~Monaca\inst{\ref{in:INAF-IAPS},\ref{in:UniRoma2}}\thanks{Corresponding author: fabio.lamonaca@inaf.it}\orcidlink{0000-0001-8916-4156}
\and Alessandro Di~Marco\inst{\ref{in:INAF-IAPS}}\orcidlink{0000-0003-0331-3259} 
\and
Francesco Coti~Zelati\inst{\ref{in:ICE},\ref{in:IEEC},\ref{in:oabr_merate}}\orcidlink{0000-0001-7611-1581}
\and 
Anna Bobrikova\inst{\ref{in:UTU}}\orcidlink{0009-0009-3183-9742}
\and
Renee M.~Ludlam\inst{\ref{in:Wayne}}\orcidlink{0000-0002-8961-939X}
\and
Juri Poutanen \inst{\ref{in:UTU}}\orcidlink{0000-0002-0983-0049}
\and
Alessio Marino\inst{\ref{in:ICE},\ref{in:IEEC}}\orcidlink{0000-0001-5674-4664}
\and
Songwei Li \inst{\ref{in:Wayne}}\orcidlink{0009-0005-8520-0144}
\and  
Fei Xie \inst{\ref{in:Guangxi},\ref{in:INAF-IAPS}}\orcidlink{0000-0002-0105-5826}
\and 
Hua~Feng \inst{\ref{in:CAS}}\orcidlink{0000-0001-7584-6236}
\and 
Chichuan~Jin \inst{\ref{in:NAO-CAS},\ref{in:UCAS},\ref{in:IFAA}}\orcidlink{0000-0002-2006-1615}
\and
Nanda~Rea \inst{\ref{in:ICE},\ref{in:IEEC}}\orcidlink{0000-0003-2177-6388}
\and
Lian~Tao \inst{\ref{in:CAS}}\orcidlink{0000-0002-2705-4338}
\and 
Weimin~Yuan \inst{\ref{in:NAO-CAS},\ref{in:UCAS}}
}

\authorrunning{F. La Monaca et al.}

\institute{
        INAF--Istituto di Astrofisica e Planetologia Spaziali, Via del Fosso del Cavaliere 100, 00133 Roma, Italy \label{in:INAF-IAPS}
        \and
        Dipartimento di Fisica, Universit\`{a} degli Studi di Roma ``Tor Vergata'', Via della Ricerca Scientifica 1, 00133 Roma, Italy \label{in:UniRoma2} 
        \and
        Institute of Space Sciences (ICE, CSIC), Campus UAB, Carrer de Can Magrans s/n, 08193 Barcelona, Spain
        \label{in:ICE}
        \and
        Institut d'Estudis Espacials de Catalunya (IEEC), 08860 Castelldefels (Barcelona), Spain
        \label{in:IEEC}
        \and
        INAF--Osservatorio Astronomico di Brera, Via Bianchi 46, 23807 Merate (LC), Italy
        \label{in:oabr_merate}
         \and 
         Department of Physics and Astronomy, 20014 University of Turku, Finland 
        \label{in:UTU}
        \and
        Department of Physics and Astronomy, Wayne State University, 666 W. Hancock St., 48201 Detroit (MI), USA \label{in:Wayne} 
        \and
        Guangxi Key Laboratory for Relativistic Astrophysics, School of Physical Science and Technology, Guangxi University, Nanning 530004, China \label{in:Guangxi}
        \and
        Key Laboratory of Particle Astrophysics, Institute of High Energy Physics, Chinese Academy of Sciences, Beijing 100049, China\label{in:CAS} 
        \and
        National Astronomical Observatories, Chinese Academy of Sciences, 20A Datun Road, Beijing 100101, China\label{in:NAO-CAS}
        \and
        School of Astronomy and Space Sciences, University of Chinese Academy of Sciences, 19A Yuquan Road, Beijing 100049, China\label{in:UCAS}
        \and
        Institute for Frontier in Astronomy and Astrophysics, Beijing Normal University, Beijing 102206, China\label{in:IFAA}
        }
        
\date{Received 11 April 2025; accepted 16 August 2025}

\abstract{This study presents an X-ray spectropolarimetric characterisation of the Z source \gx during the normal branch (NB) and compares it with that obtained for the horizontal branch (HB), using IXPE, NICER, and NuSTAR observations. The analysis reveals significant polarisation, with polarisation degrees of ${\sim}1.4$\% in the NB and ${\sim}3.7$\% in the HB, indicating a notable decrease in polarisation when transitioning from the HB to the NB. The polarisation angles show a consistent trend across the states. Spectropolarimetric analysis favours a dependence of the polarisation on the energy. The Comptonised component shows similar polarisation in both the HB and NB and is higher than the theoretical expectation for a boundary or spreading layer. This suggests a contribution from the wind or the presence of an extended accretion disc corona (ADC) to enhance the polarisation. The results obtained here highlight the importance of using polarimetric data to better understand the accretion mechanisms and the geometry of this class of sources, providing insights into the nature of the accretion flow and the interplay between different spectral components. Overall, the findings advance our understanding of the physical processes governing accretion in low-mass X-ray binaries.}


\maketitle
%

\section{Introduction}\label{sec:intro}

Low-mass X-ray binaries (LMXBs) that host accreting weakly magnetised neutron stars (WMNSs) are systems in which the donor star, with a mass that is typically smaller than one solar mass, transfers matter to the neutron star (NS) via Roche-lobe overflow, forming an accretion disc. A disc coplanar boundary layer (BL) can form as the accretion flow slows from the high velocity of the disc to the lower velocity at the NS surface \citep{Shakura88, Popham01}. Furthermore, accreting matter can spread across the NS surface, forming a spreading layer (SL) that can extend vertically with respect to the disc up to high latitudes on the NS surface \citep{inogamov1999, suleimanov2006}. The WMNSs are classified, on the basis of the shape they track in the X-ray colour-colour diagram (CCD) and the hardness-intensity diagram (HID) \citep[see, e.g.,][]{vanderklis89,hasinger89}, into Z and Atoll sources. The Z sources, which show luminosities near the Eddington limit, follow a Z shape divided into three distinct branches: horizontal (HB), normal (NB), and flaring (FB); regions connecting the branches are named the hard apex (HA) and the soft apex (SA) \citep{Church12, Motta19}. These different states correspond to different accretion rates ($\Dot{M}$), which are characterised by the different hardnesses of their spectra: the FB spectrum is softer with higher $\Dot{M}$, while the HB spectrum is harder with lower $\Dot{M}$. Atolls are fainter compared to Z sources and display a hard state named Island and a soft state named Banana. For general reviews of WMNSs see, e.g., \cite{Done07}, \cite{Bahramian23}, \cite{DiSalvo24}.

Six persistent Z sources are known so far; traditionally, they are further divided into Sco-like and Cyg-like ones \citep{Kuulkers94, kuulkers97}. \mbox{Sco~X-1}, \mbox{GX~349$+$2}, and \mbox{GX~17$+$2} belong to the first group, while \mbox{Cyg~X-2}, \mbox{GX~5$-$1}, and \gx belong to the second one. Cyg-like and Sco-like sources have different Z tracks. Cyg-like sources have a full Z track and a weak flaring state, while Sco-like sources have a Z track with an HB that is difficult to spot and a higher-flux flaring state. In addition, Cyg-like sources are supposed to have a higher inclination or a higher magnetic field than Sco-like sources \citep{Psaltis95, kuulkers97, Church12}.

One of the open questions in understanding WMNSs is the geometry of their accretion flow and the mechanism at the basis of it; in particular, the shape of the region between the inner disc and the NS surface and the contributions of each emitting region (BL, SL, inner disc, and NS surface) to their spectra are still a matter of debate \citep[see, e.g.,][and references therein]{Ludlam24}. X-ray spectra of WMNSs have a continuum in which a soft and a harder component can be identified. The soft one can be described by a multi-colour or single-temperature black body that can be associated with the emission from the inner accretion disc or the NS surface or both of them, while the harder one can be described by a Comptonised continuum that can originate from the BL and/or SL. In addition, the Comptonised medium can also be an optically thin plasma located above and below the disc, named the accretion disc corona (ADC) \citep[see, e.g.,][]{White82, Parmar88, Miller00}. A typical feature in the WMNSs spectra is the presence of a fluorescence iron line at ${\sim}6.4$\,keV; this is the main feature of a reflection spectrum due to the emission of reprocessed radiation by the inner accretion disc. Studying the reflection spectrum gives an insight into the inner region near the NS \citep[see, e.g.,][]{Bhattacharyya07, Cackett08, Miller13, Ludlam17, Ludlam22, Ludlam24}. Spectral properties, supported by timing studies \citep{gilfanov2003, Revnivtsev06, Revnivtsev13}, associate quasi-periodic oscillations with a hard stable component, favouring the presence of a Comptonising medium between the inner disc and the NS surface; thus, these results agree with predictions for BL and/or SL reported in the so-called eastern model scenario \citep{Mitsuda84, Mitsuda89}.

Recently, thanks to the Imaging X-ray Polarimetry Explorer \citep[\ixpe;][]{Soffitta2021, Weisskopf2022}, a new diagnostic tool for studying the geometry and accretion mechanism of WMNSs has become available: X-ray polarisation. It adds two new observables, the polarisation angle (PA) and the polarisation degree (PD), to the existing information obtained by the spectral and timing properties. Since its launch in 2021, IXPE has observed all persistent Z sources, measuring a higher PD in the HB compared to the NB, SA, and FB \citep[see, e.g.,][]{DiMarco25a}. Similar values for the PDs of the HB and NB were also measured for the transient \mbox{XTE J1701$-$462} \citep{Cocchi23}, which showed both Z and atoll peculiarities; this source showed hints of variation in the PA with time in the NB \citep{DiMarco25a, Zhao25}. Other WMNSs showed variations of the PA with the source state and time, such as \mbox{Cir~X-1} \citep{Rankin24} and \mbox{GX~13$+$1} \citep{Bobrikova24a, Bobrikova24b, DiMarco25b}. To study the geometry of WMNSs, it is interesting to correlate the measured PA with the radio jet's direction, which is almost perpendicular to the disc plane. Unfortunately, the direction of the radio jet was measured only for a few WMNSs \citep{Fomalont01, Fomalont01b, Spencer13}. \ixpe observed a PA aligned with the radio jet for \mbox{Cyg X-2} \citep{Farinelli23}, while it measured a PA rotated ${\sim}46\degr$ for \mbox{Sco X-1} \citep{LaMonaca24a} and a PA not aligned with the radio jet for \mbox{Cir~X-1}, where a rotation with time and hardness was reported \citep{Rankin24}. A review of IXPE results on WMNSs can be found in, for example, \citet{Ursini24WMNSs_IXPE} and \citet{DiMarco25a}.

\ixpe observed \gx (4U~1642$-$45) twice, with the first observation covering the HB \citep{LaMonaca24gx, Bhargava24} and the second covering the NB, as is reported here and in \citet{Bhargava24b}, in which a different approach to the analysis and data interpretation was adopted. For the HB, \ixpe reported a highly significant detection of polarisation with $\textrm{PD}=4.3\%\pm0.4\%$ and $\textrm{PA}=36\degr\pm3\degr$. The spectropolarimetric analysis, based on a broadband modelling of the \nicer and \nustar quasi-simultaneous observations, revealed a significant polarisation for both the soft and the hard components. Notably, their PAs differ ${\sim}40\degr$, even though, based on the geometry and optical depth of the BL and/or SL, they would typically be expected to be nearly parallel or orthogonal \citep[see, e.g.,][]{st85}. This result may be compatible with a misalignment of the NS axis with respect to the accretion disc axis. The same conclusions were obtained for \mbox{Cir~X-1} \citep{Rankin24} and \mbox{GX~13$+$1} \citep{Bobrikova24a}. Recently, \cite{LaMonaca24gx}, \cite{Li25}, and \cite{Ludlam25} derived the inclination of the source, studying the reflection spectrum with \texttt{relxillNS} \citep{Garcia22}, obtaining 34--40~deg, in line with the previous results of \cite{Dai09}, \cite{Cackett10}, and \cite{Miller16} and the estimation by radio observation of \cite{Fender00}. This inclination is not expected for a Cyg-like source \citep{Kuulkers96}, and the measured PD for the disc and the Comptonised components from theoretical polarimetric models \citep{st85, Loktev22, Farinelli24, Bobrikova25}, also favours an inclination of \gx that is higher than 60\degr \citep{LaMonaca24gx}. Furthermore, for \gx previous studies in literature have reported evidence of wind \citep{Miller16} and of the presence of an ADC \citep{Church06, Ludlam25}.

This paper presents the results of the second \ixpe co-ordinated observational campaign, and it is structured as follows. The observations and data reduction are presented in Sect.~\ref{sec:observations}. The polarimetric and spectropolarimetric analyses are presented in Sect.~\ref{sec:data_results}. The discussion and conclusion are given in Sect.~\ref{sec:discussion}.

\section{Observations and data reduction}\label{sec:observations}

\begin{table*}[!hbt]
\centering
\caption{List of telescopes and/or instruments with the respective observation IDs and the exposure times of each observation.}
\label{tab:exposure}
\begin{tabular}{llcccc}
\hline \hline
& Obs ID & Start (UTC) & Stop (UTC)  &Telescope& Exp. time (s)\\
 \hline\

\ixpe & 03009901 & 2024-08-12 07:25 & 2024-08-16 06:46 & DU 1 & 191627\\
 & & & & DU 2 & 191867 \\
 & & & & DU 3 & 191851\\
\hline 
\nicer  & 7705010101 & 2024-08-12 06:18 & 2024-08-12 21:59 & & 3064\\
        & 7705010102 & 2024-08-13 00:52 & 2024-08-13 21:12 & & 4141\\
        & 7705010103 & 2024-08-14 00:06 & 2024-08-14 23:31 & & 4012\\
        & 7705010104 & 2024-08-15 02:24 & 2024-08-15 22:45 & & 2736\\
        & 7705010105 & 2024-08-16 00:04 & 2024-08-16 04:56 & & 1370\\

\hline
\ep     & 06800000043 & 2024-08-12 13:42 & 2024-08-12 14:43 & WXT/CMOS13 & 763\\
        & 11900004570 & 2024-08-15 02:29 & 2024-08-15 02:58 & WXT/CMOS24 & 1505\\
        & 08500000140 & 2024-08-16 02:33 & 2024-08-16 04:05 & WXT/CMOS2  & 2675\\        
\hline
\nustar & 30302030006 &2018-08-17 07:01 & 2018-08-17 19:41 & FPMA &  14890\\
 & & & & FPMB & 15355\\
\hline
\end{tabular}
\end{table*}

The X-ray data from \ixpe, \nicer, and \nustar observatories are publicly available in NASA's High-Energy Astrophysics Science Archive Research Center (HEASARC). The light curves and HID presented in this paper are obtained using \textsc{stingray} \citep{stingray2,stingray1,Bachetti24}. For the X-ray observatories, except in cases where it is explicitly written, the data were extracted using standard pipelines and \textsc{ftools} included in HEASoft version 6.34 \citep{heasoft}. Spectral and spectropolarimetric analyses were performed using \textsc{xspec} \citep{Arnaud96}.

\subsection{\ixpe}\label{subsec:ixpe}

\ixpe is a NASA-ASI mission launched in 2021; it enabled the X-ray linear polarisation in the band \mbox{2--8\,keV} to be measured for the first time for almost all classes of astrophysical sources. The satellite comprises three identical telescopes. Each telescope consists of a multi-mirror array with a detector unit (DU) that houses a photoelectric polarimeter in the focal plane. For the satellite performance, the description and the working principle of the on-board instrument, see \cite{Soffitta2021}, \cite{Baldini21}, \cite{Weisskopf2022}, \cite{DiMarco22b}, and references therein.

\ixpe observed \gx for the second time from UTC 07:25 on August 12 2024  to UTC 06:46 on August 16 2024; the \ixpe observation ID and the exposure time for each DU are reported in Table~\ref{tab:exposure}. We processed the data with the \textsc{ixpeobssim} package version 31.0.3 \citep{Baldini22}. This software, developed within the \ixpe collaboration, allows for model-independent polarimetric analysis using the \texttt{xpbin} tool with the \texttt{pcube} algorithm based on the \citet{Kislat15} approach. The \ixpe data were binned to have a minimum of 30 counts per bin for the Stokes $I$ spectrum, and a constant binning of 200\,eV for the Stokes $Q$ and $U$ ones. The \ixpe data were analysed using the CALDB response matrices 20240701 released with \textsc{ixpeobssim} on February 28 2024. The source was selected in a 100\arcsec\ radius circular region centred on the source position using \textsc{SAOImageDS9}.\footnote{\href{https://sites.google.com/cfa.harvard.edu/saoimageds9}{https://sites.google.com/cfa.harvard.edu/saoimageds9}} Due to the relatively high flux of the source and following the \ixpe prescription reported in \cite{DiMarco23a}, the polarimetric analysis was performed without subtracting the background. 

\subsection{\nicer}\label{subsec:NICER}
The Neutron Star Interior Composition Explorer (\nicer) is an X-ray instrument operating on board the International Space Station since 2017. It consists of 56 co-aligned concentrator X-ray optics focusing the X-rays on silicon drift detectors, and it works in the 0.2--12\,keV energy band. A detailed description of the instrument can be found in \cite{Gendreau16}. 

\nicer performed several short observations of \gx collected in five simultaneous observation IDs during the second \ixpe observation reported here. Table~\ref{tab:exposure} shows the \nicer observation IDs and their start and stop dates with total exposure times. \nicer data were processed with the \nicer Data Analysis Software v012a and with the CALDB version \textsc{XTI20240206}. The \nicer spectra were extracted with the \texttt{nicerl3-spec} task, applying the \nicer-recommended systematic error vector to the data and optimally binned to have at least 30 counts per bin. The background was estimated using the \textsc{SCORPEON} model; both the full model and the file version were used, and the results were compatible; thus, in the final analysis here reported, we used the file version.   

\subsection{\nustar}\label{subsec:NuSTAR}
The Nuclear Spectroscopic Telescope Array \citep[\nustar;][]{Harrison2013} is an X-ray observatory operating in the 3--79\,keV energy band. It comprises two identical telescopes (FPMA and FPMB), each coupling a Wolter-I X-ray optics and a CdZnTe pixel detector. 
During this new \ixpe observation of \gx, there were no strictly simultaneous pointings by \nustar. Therefore, as is described in Sect.~\ref{sec:data_results}, after identifying the source state in a HID with \ixpe and \nicer, we selected in the \nustar archive an observation of \gx in the same state, reported in Table~\ref{tab:exposure}, aiming to obtain a broadband spectral model. 

The \nustar data were processed with the standard Data Analysis Software (NuSTARDAS) v2.1.2 with the CALDB version 20241015. As is prescribed for bright sources,\footnote{\href{https://heasarc.gsfc.nasa.gov/docs/nustar/analysis/}{https://heasarc.gsfc.nasa.gov/docs/nustar/analysis/}} we set the statusexpr="STATUS==b0000xxx00xxxx000" keyword in the \texttt{NuPIPELINE}. 
The source signal was extracted in a circular region with a radius of  120\arcsec\ at the position of the source, and the background region is selected in an off-centre sourceless circular region with the same radius. The \nustar spectra were binned to have at least 30 counts per bin.

\subsection{\ep}\label{subsec:EP}

The Einstein Probe \citep[EP;][]{Yuan2022,Yuan2025} is a mission launched in January 2024 and dedicated to time-domain X-ray astronomy. EP features a wide-field X-ray telescope (WXT) that employs innovative lobster-eye micro-pore optics to monitor the soft X-ray sky (0.5--4\,keV) over an instantaneous field of view of $\simeq$3600\,deg$^2$ (see also \citealt{Cheng2025}). Complementing this, the follow-up X-ray telescope (FXT), based on a Wolter-I design, enables deeper spectral and timing studies in the 0.3--10\,keV band. 

EP-WXT covered the position of \gx three times during our campaign (see Table~\ref{tab:exposure} for details). Starting from the cleaned event files produced by the processing pipeline available within the WXT Data Analysis Software (\texttt{WXTDAS}), the source photons were extracted from a circular region centred on the source position with a radius of 9\arcmin\ (corresponding to $\simeq$90\% of the encircled energy fraction of the instrument’s point spread function). The background region was defined using a circle with the same radius, located nearby. Background-subtracted light curves were extracted in the 0.5--4, 0.5--2, and 2--4\,keV energy bands and binned at 600\,s.

\section{Data analysis and results}\label{sec:data_results}

\begin{figure}[!h]
   \centering
   \includegraphics[width=0.9\columnwidth]{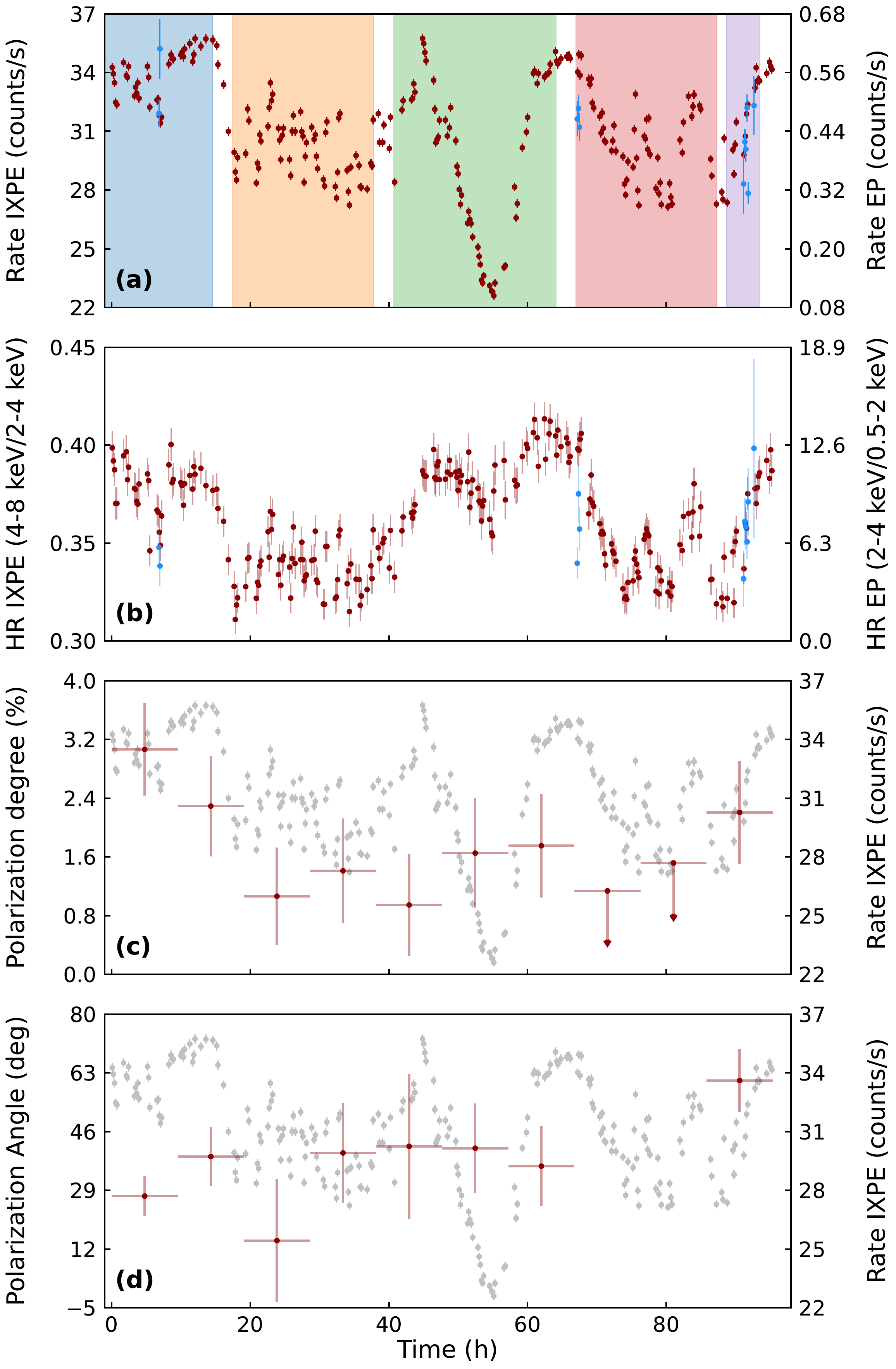}
   \caption{Light curves, HRs, and polarisation properties of \gx during the second \ixpe observation. \emph{(a)}: \ixpe light curve binned in 600\,s intervals overlapped with the EP-WXT light curve (dodger blue points). The highlighted coloured intervals correspond to the \nicer simultaneous observations reported in Table~\ref{tab:exposure} and Fig.~\ref{fig:NICER_NuSTAR_HID}. \emph{(b)}: \ixpe HR binned in 600\,s  obtained as the ratio of the \ixpe counting rates in the 4--8 and 2--4\,keV energy bands.  \emph{(c)} and \emph{(d)}: Time-resolved PD and PA in equal time bins of about 9.5\,h. The errors are reported at 68\%~CL, while the upper limits are at 90\%~CL.\label{fig:IXPE_LC_HR_PD_PA}}
\end{figure}

The light curve of this new \ixpe observation is reported in Fig.~\ref{fig:IXPE_LC_HR_PD_PA}a, while Fig.~\ref{fig:IXPE_LC_HR_PD_PA}b reports the corresponding hardness ratio (HR). The \ixpe HR has been obtained as the ratio of the \ixpe counting rates in the 4--8 and 2--4\,keV energy bands, consistent with the definition adopted by \cite{LaMonaca24gx}. The light curve shows variations, with a significant drop in flux at about 50 hours since the beginning of the observation. This flux variation corresponds to higher HR values. Figure~\ref{fig:IXPE_LC_HR_PD_PA} also reports the light curves and HR from the EP-WXT observations overlaid on those from the \ixpe one. Although the short durations of the EP-WXT observations do not allow us to fully track the source state variations, the overall trends in flux and HR are in agreement with those from \ixpe during the same time intervals.

\begin{figure}
   \centering
   \includegraphics[width=0.9\columnwidth]{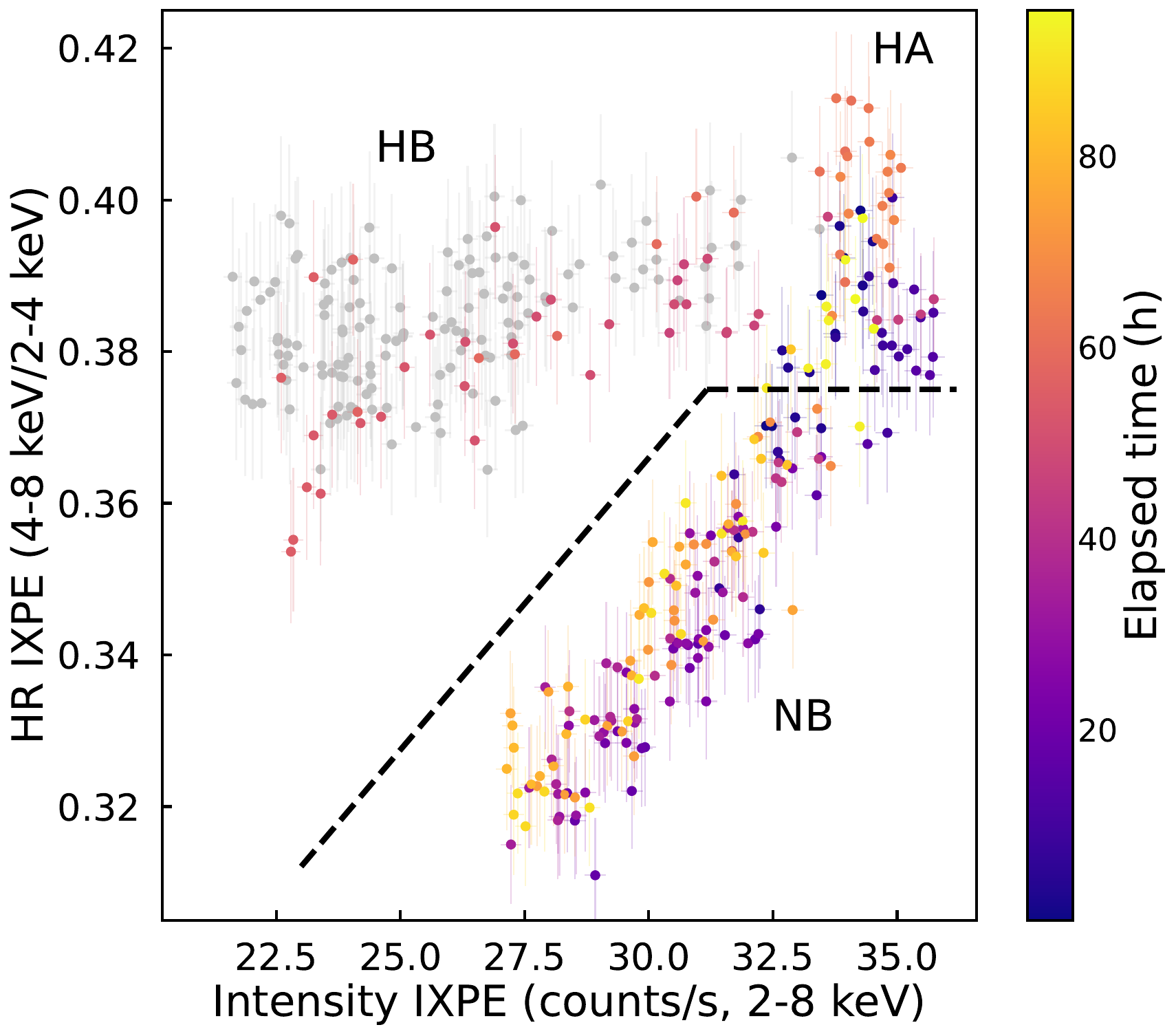}
   \caption{\ixpe HID of the second \ixpe observation of \gx in 600\,s time bins. The grey points represent the first \ixpe observation, while the coloured ones, from dark blue to yellow, report the elapsed time in hours since the start of the second \ixpe observation. The dotted black line defines the region we used to select the NB. \label{fig:IXPE_HID}}
\end{figure}

To identify the state of the source during this observation and possible state variations, we constructed the \ixpe HID, shown in Fig.~\ref{fig:IXPE_HID}; data points are colour-coded with the elapsed time in hours since the start of the observation. The energy bands we adopted are the same as those used by \citet{LaMonaca24gx} for easier comparison with the first \ixpe observation, whose points are shown in grey. Comparing the HID from the two observations, we note that at the beginning of this second observation, \gx was in the HA, then moved to the NB, came back to the HA, moved to the HB (coinciding with a flux decrease in the middle of the observation) and in the last part, moved back to the NB. Thus, this new observation covered mainly the NB, with short excursions into the HB and HA.

\begin{figure}[!h]
   \centering
   \includegraphics[width=0.8\columnwidth]{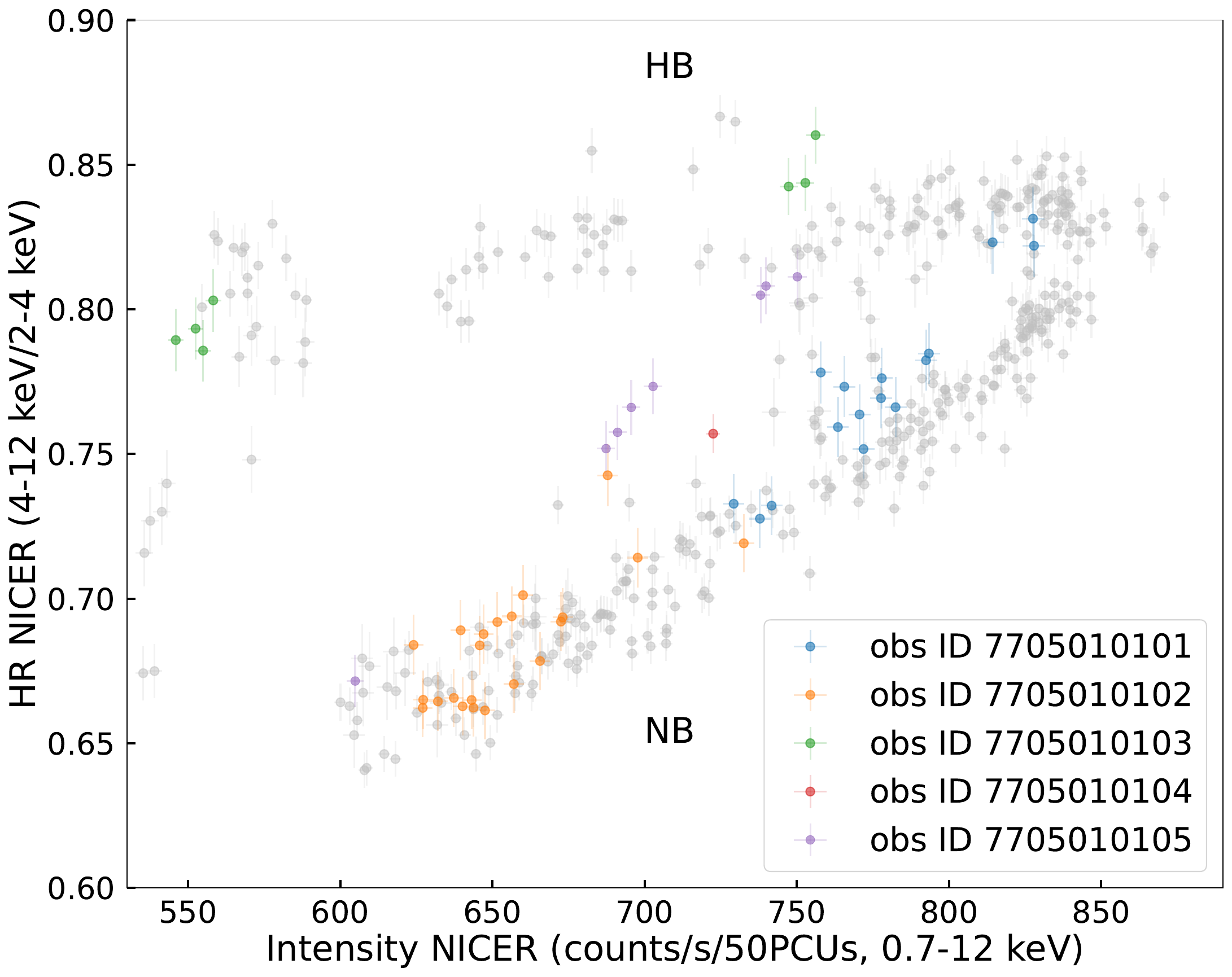}
   \includegraphics[width=0.8\columnwidth]{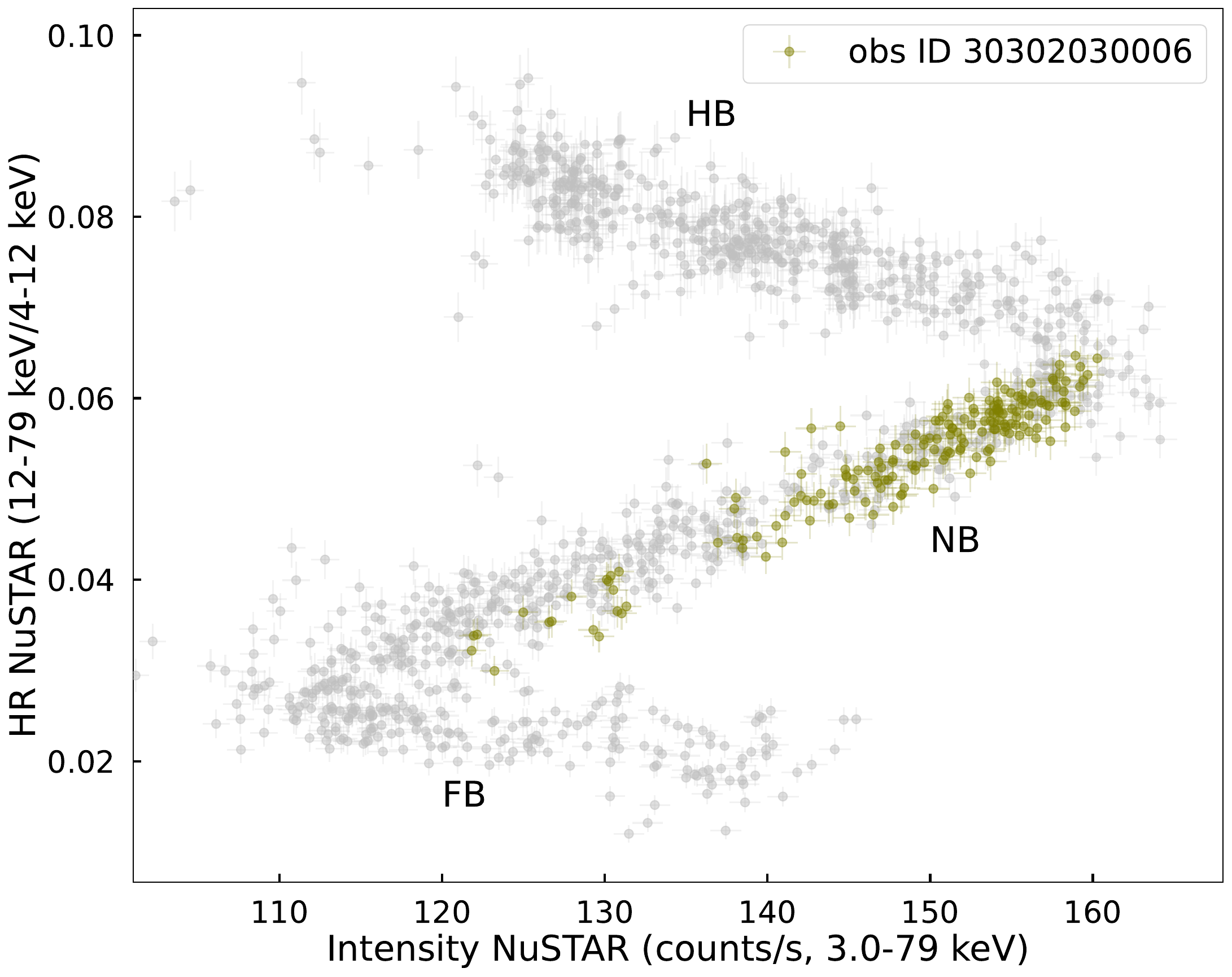} 
   \caption{HIDs of \gx in 150\,s time bin.  
   \emph{Top panel}: HID of \gx from the \nicer data archive (grey points). The coloured points represent the five \nicer observations performed during the IXPE one and also reported in Table~\ref{tab:exposure} and Fig.~\ref{fig:IXPE_LC_HR_PD_PA}-\emph{top} as shaded coloured regions in the \ixpe light curve. \emph{Bottom panel}: HID of \gx from the \nustar archival data  (grey points). The olive green points represent the \nustar observation selected to perform the spectral and spectropolarimetric analysis. \label{fig:NICER_NuSTAR_HID}}
\end{figure}

The \gx state variations are confirmed from the \nicer observations performed jointly with the \ixpe one, reported in Fig.~\ref{fig:NICER_NuSTAR_HID}-\emph{top}. In fact, the \nicer HID of the simultaneous observations (coloured points) compared to all the archived \nicer observations of \gx (grey points) shows that \gx was mainly in NB with short periods in HA and HB. We selected the Observation ID 7705010102 for the spectral analysis.

Unfortunately, no strictly simultaneous \nustar observations were available for this second \ixpe observation. \nustar Observation ID 30901012002 was performed 5 days after this \ixpe observation; \gx, during this \nustar observation, was in NB, SA, and FB. When selecting only the NB, the exposure time becomes too low, making it preferable to use older observations to obtain a broadband spectral model of the NB. For this purpose, we used archived \nustar observations to build the HID reported in Fig.~\ref{fig:NICER_NuSTAR_HID}-\emph{bottom} and selected the observation ID 30302030006, which covered \gx in the NB, to have a suitable observation to use for proper spectral modelling. 

\subsection{Polarimetric analysis}

In this subsection, we report the results obtained by the polarimetric analysis performed using the \texttt{pcube} algorithm in \textsc{ixpeobssim} software \citep{Baldini22}. Firstly, we investigate possible polarimetric variations with time. Figures~\ref{fig:IXPE_LC_HR_PD_PA}c and d show PD and PA obtained by dividing the \ixpe observation into ten time bins of $\sim$9.5\,h each; the chosen time binning is a trade-off between suitable time resolution and polarimetric sensitivity. The achieved sensitivity does not show any significant variation in PD and PA with either time or source flux, except for possible variations related to the different Z branches (see below). 

Given the changes in the source's state during this observation, with short periods of HA and HB, we focused our analysis on the polarimetric properties in the NB. To separate the period when the source is in the NB from the one in which it is in the HB and HA, we set a threshold on the \ixpe intensity and HR (region below the black line of Fig.~\ref{fig:IXPE_HID}). 

The X-ray polarisation measured in the nominal 2--8\,keV energy band, after selecting the NB, is $\textrm{PD} = 1.4\%\pm0.3\%$ with a $\textrm{PA} = 37\degr\pm5\degr$ at 68\% confidence level (CL). Therefore, we have a detection of polarisation at $\sim$5.6$\sigma$ CL, corresponding to a probability of obtaining this polarisation in the case of an unpolarised source of $1.1\times 10^{-7}$.

\begin{figure}[!h]
   \centering
   \includegraphics[width=0.85\columnwidth]{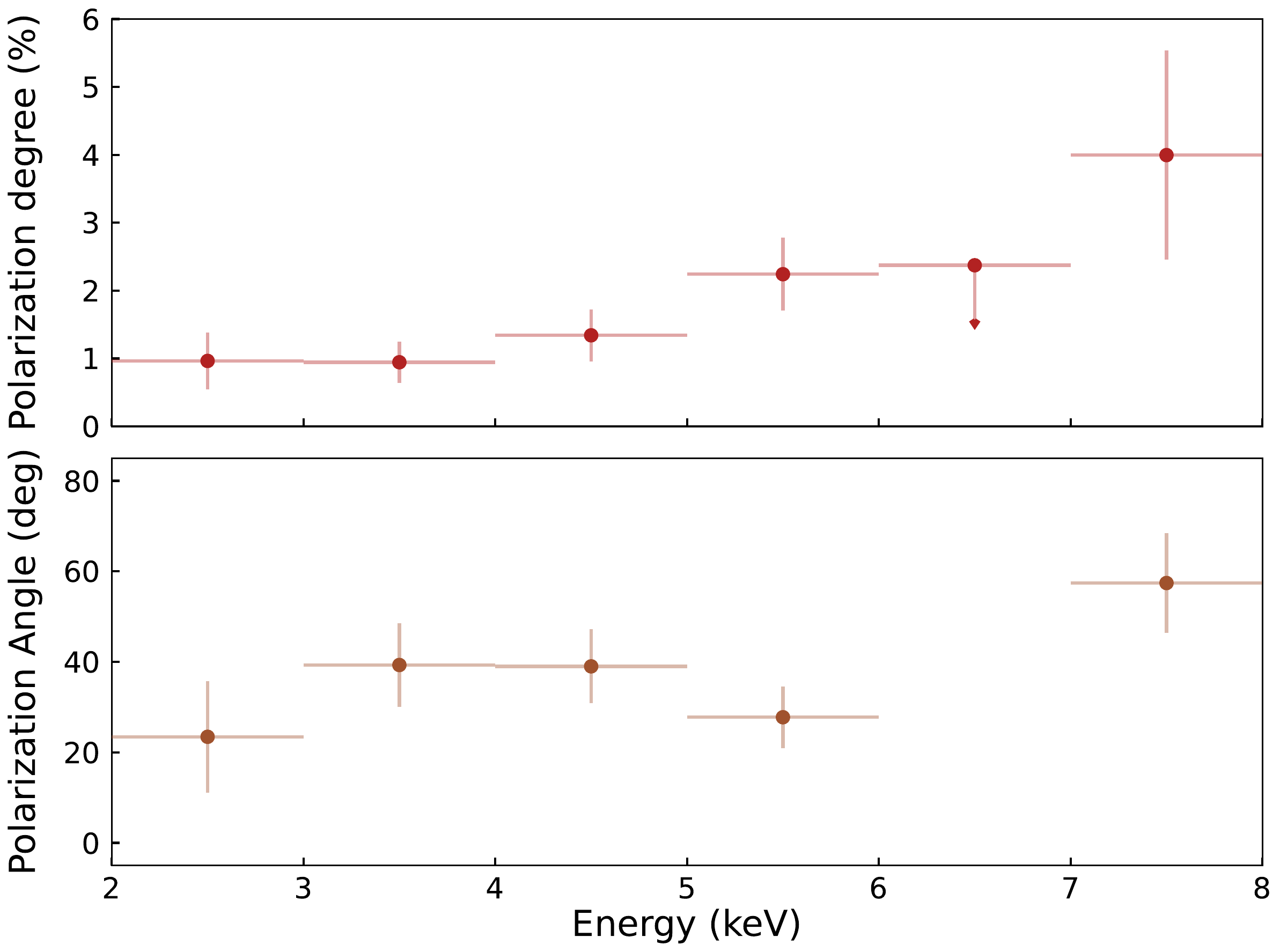}
   \caption{Energy-dependent polarimetric properties of \gx in the NS. PD (top panel) and PA (bottom panel) as a function of the energy in 1\,keV energy bins. The errors are reported at 68\%~CL, while the upper limit is at 90\%~CL, see text.\label{fig:PD_PAvsEnergy}}
\end{figure}

We performed an energy-resolved analysis of the polarisation of the NB state of \gx, following the same approach reported in \citet{LaMonaca24gx} but dividing the nominal \ixpe energy band into 1\,keV-wide energy bins to enhance the polarimetric significance. The results are shown in Fig.~\ref{fig:PD_PAvsEnergy} and Table~\ref{tab:PD_PAvsEnergy}, where only an upper limit at 90\%~CL is reported for the 6--7\,keV bin, given its lower significance (1.5$\sigma$), while the others show a significance higher than 99\%~CL. 
\begin{table}[!h]
\centering
\caption{Energy-resolved polarisation  of \gx in the NB.}
\label{tab:PD_PAvsEnergy}
\begin{tabular}{ccc}
\hline \hline
Energy Bin & PD & PA\\
(keV) & (\%) & (deg)\\
 \hline\
2.0--3.0 & $1.0 \pm 0.4$ & $23 \pm 12$ \\ 
3.0--4.0 & $0.9 \pm 0.3$ & $39 \pm 9$ \\ 
4.0--5.0 & $1.3 \pm 0.4$ & $39 \pm 8$ \\ 
5.0--6.0 & $2.2 \pm 0.5$ & $28 \pm 7$ \\ 
6.0--7.0 & $1.2 \pm 0.8$ & $31 \pm 19$ \\ 
7.0--8.0 & $4.0 \pm 1.5$ & $57 \pm 11$ \\ 
\hline
2.0--8.0 & $1.4\pm0.3$  &  $37\pm5$  \\
\hline
\end{tabular}
\tablefoot{The polarisation values were obtained with the \texttt{pcube} algorithm in the 2--8\,keV energy band for each 1\,keV-wide energy bin. The errors are reported at 68\%~CL.}
\end{table}

\begin{figure}[!h]
   \centering
\includegraphics[width=0.8\columnwidth]{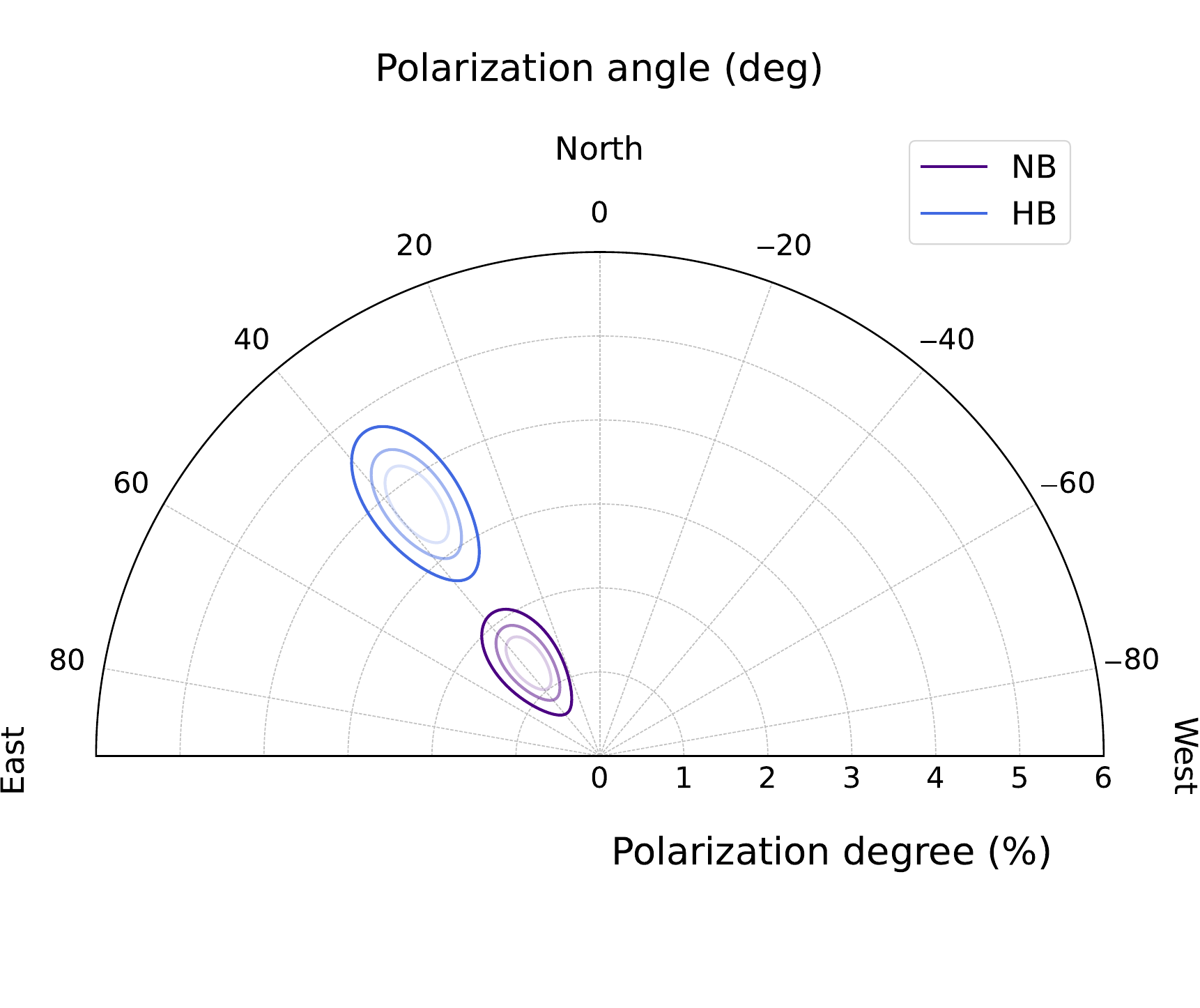}
   \caption{Polarisation in the HB (blue) and in the NB (purple) of \gx. The contours represent the allowed regions at 68\%, 90\%, and 99\% CL. \label{fig:hb_nb}}
\end{figure}

We tried to select only HA or HB, or consider both (HA+HB), obtaining $\textrm{PD}{<}6\%$, $\textrm{PD}{<}3$\%, and $\textrm{PD}{<}3$\% at 99\% CL, respectively. Thus, while the analysis of the more recent observation yields values compatible with those obtained in \cite{LaMonaca24gx}, the significance of HB is much lower, mainly due to the shorter exposure time covering \gx in HB+HA. Combining the new data of this observation in HB with the previous ones, we obtain a marginal improvement in the significance of the polarisation: $\textrm{PD} = 3.7\%\pm0.4\%$ with a $\textrm{PA} = 36\degr\pm2\degr$ (errors are at 68\% CL). In Fig.~\ref{fig:hb_nb}, we compare the polarisation in the NB with that obtained for the HB by combining the two observations.
It is possible to notice the higher PD in HB with respect to NB, while the PA remains compatible.

\subsection{Spectropolarimetric analysis} \label{sec:spec}

In \cite{LaMonaca24gx}, the spectropolarimetric analysis for the HB of \gx is reported; as was previously stated, this new observation could give just a marginal improvement in the significance of the measured polarisation in HB, which makes a new spectropolarimetric study for this branch not useful. Thus, we performed a spectropolarimetric analysis only for the NB of \gx, to compare if there is any difference between the two branches. 

Considering the narrow energy band and the poor spectral resolution of \ixpe, we need to constrain our spectral model with a joint \nicer + \nustar + \ixpe broadband spectral analysis. After selecting observations and data, as was previously reported, we performed a joint fit using \nicer in (2--12\,keV), \nustar in (3--30\,keV), and \ixpe (2--8\,keV). The same approach has been adopted in previous papers for this class of sources \cite[see, e.g.,][]{LaMonaca24gx}. To take into account instrumental effects, we kept the gain free to vary in the fit and added to the \nicer spectral model a line at 1.74\,keV and an edge at 1.84\,keV. Although below our \nicer energy threshold, the tail of these features also has an impact above 2\,keV.

\begin{figure}
\centering
\includegraphics[width=0.95\linewidth]{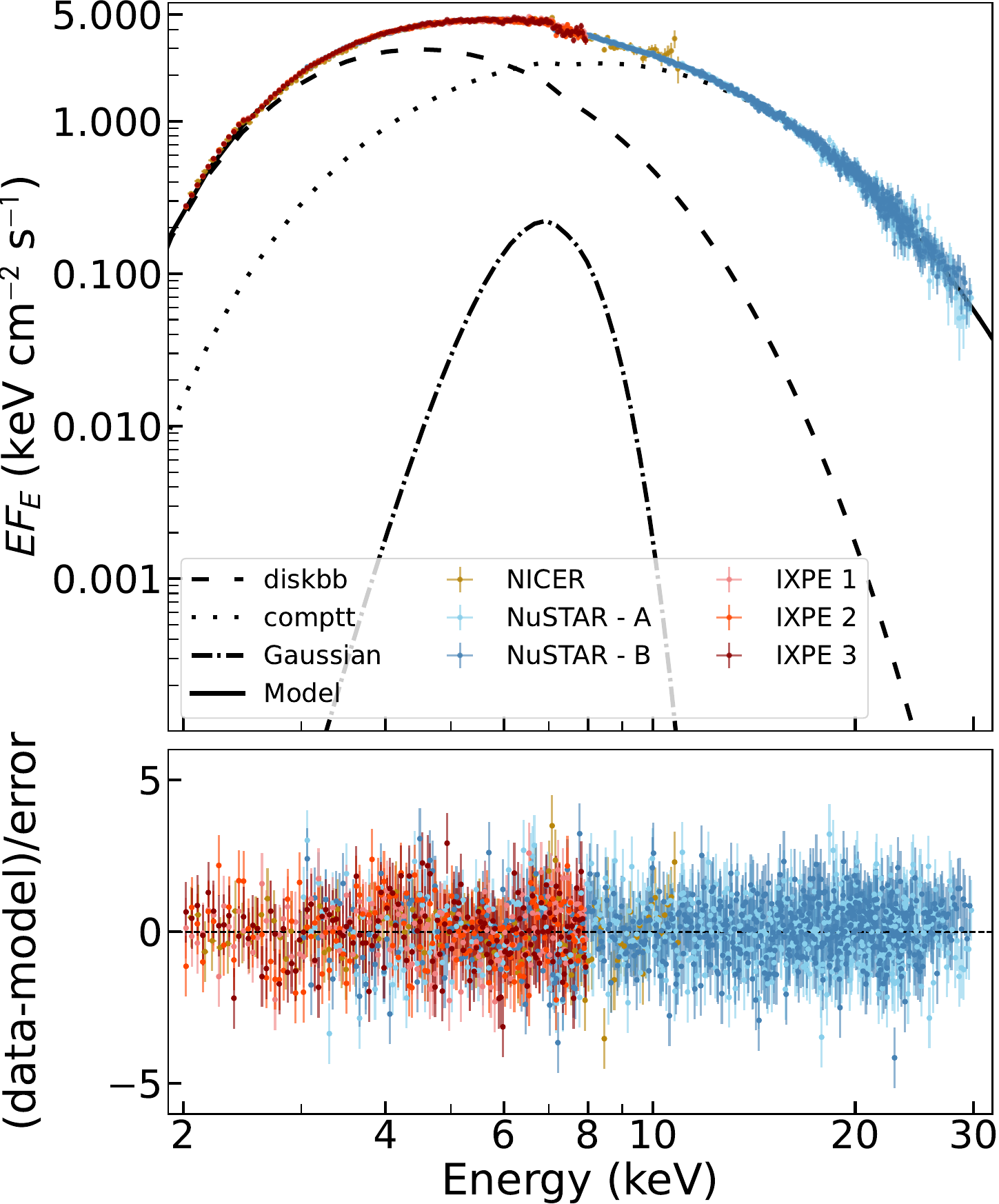}
\caption{Spectral energy distribution of \gx in $EF_E$ representation for the simultaneous fit using Model~A of \nicer (dark yellow points), \ixpe (red points), and \nustar (blue points) observations. The different spectral model components are reported in black lines for \texttt{diskbb} (dashed), \texttt{bbodyrad} (dotted), and \texttt{Gaussian}  (dash-dotted). The \emph{bottom panel} shows the residuals between the data and the best-fit Model~A. The best-fit parameters are reported in Table~\ref{tab:spectrum}.}
\label{fig:bestfitA}
\end{figure}

To model the continuum with \textsc{xspec}, for the interstellar absorption model \texttt{tbabs} we set the element abundances at the \texttt{wilm} values \citep{2000ApJ...542..914W}, and we used the \texttt{diskbb} model for the soft spectral component representing the accretion disc and a Comptonisation model \texttt{comptt} for the harder one associated with the BL and/or SL. We also tested other possible models for the Comptonisation, with different seed photons, but no improvement in reduced $\chi^2$ is observed; thus, we used \texttt{comptt} coherently with the fit used in HB by \cite{LaMonaca24gx}. The \texttt{comptt} model was used as in \cite{LaMonaca24gx}, assuming the \texttt{redshift} parameter at 0 because \gx is a galactic source, and the \texttt{approx} at 1, meaning a Comptonising medium such as a slab (similarly to a BL); we also tested the scenario with a spherical Comptonising medium, such as SL or a vertically extended BL, but we found no improvement in the $\chi^2$ and also the other parameters in the continuum are almost the same, with the exception of $\tau_{sphere}\sim2\tau_{slab}$, as has also been reported in the literature \citep{Farinelli23}. An excess in the residuals at ${\sim}6.7$\,keV is observed with respect to the continuum; this is associated with the iron fluorescence line due to reflection from the inner disc. To compensate for this excess, we added a Gaussian to the model, noted as Model~A: \texttt{tbabs*(diskbb+comptt+gauss)}. Adding the Gaussian line produces an improvement in the $\chi^2/\textrm{d.o.f.}$ passing from 2735/1693 to 1793/1690. A cross-normalisation constant is included for each detector and/or telescope. The best-fit parameters for Model~A are reported in Table~\ref{tab:spectrum}. In the following, errors are reported at 90\% CL if not stated otherwise. The spectral energy distribution with residuals is shown in Fig.~\ref{fig:bestfitA}. The cross calibration gives a gain slope of ${\sim}0.96$\,keV$^{-1}$, ${\sim}1.00$\,keV$^{-1}$ and ${\sim}0.99$\,keV$^{-1}$ for \nicer, \nustar, and \ixpe, respectively; while the gain offsets are ${\sim}80$\,eV, ${\sim}90$\,eV, and ${\sim}20$\,eV. Those values show no strong variation when the different adopted spectral models are applied.

\begin{table*}[h]
\centering
\caption{Best-fit parameters of the broadband joint spectral analysis for \gx in the NB obtained by \nicer, \nustar, and \ixpe.}
\label{tab:spectrum}
\begin{tabular}{lrcc}
\hline \hline
Model & Parameter (units) & Model~A & Model~B \\ \hline
\texttt{TBabs} & $N_{\rm H}$ ($10^{22}$ cm$^{-2}$) & $9.76_{-0.07}^{+0.04}$ & $ 9.06_{-0.04}^{+0.07}$ \\ \hline
\texttt{diskbb} & $kT_{\rm in}$ (keV) & $1.303_{-0.019}^{+0.002}$ & $1.390\pm0.014$\\
                & norm ($[R_{\rm in}/D_{10}]^2\cos\theta$) & $245.8_{-0.3}^{+6.5}$  & $181_{-6}^{+7}$\\ \hline
\texttt{comptt} & $T_0$ (keV) & $1.706_{-0.032}^{+0.004}$ & $1.72\pm0.03$\\
                & $kT$ (keV) & $3.199_{-0.042}^{+0.004}$ & $3.18\pm0.07$\\
                & $\tau$   & $3.564_{-0.010}^{+0.013}$ & $3.1_{-0.2}^{+0.3}$\\
                & norm     & $ 0.3069_{-0.0144}^{+0.0005}$ & $0.267_{-0.013}^{+0.012}$\\ \hline
\texttt{Gaussian} & $E$ (keV) & $6.62_{-0.04}^{+0.05}$ & --\\
                & $\sigma$ (keV) & $1.00_{-0.05}^{+0.03}$ & --\\
                & norm (photon~cm$^{-2}$~s$^{-1}$) & $0.013\pm0.001$ & --\\
                & Equivalent width (eV) & $11.95_{-0.03}^{+0.01}$ & --\\ \hline
\texttt{relxillNS} & Emissivity & -- & $2.30_{-0.13}^{+0.20}$ \\
                & $R_{\rm in}$ ({ISCO}) & -- & $1.2\pm0.2$ \\
                & $R_{\rm out}$ ($GM/c^2$) & -- & [1000] \\
                & Inclination (deg) & -- & $33_{-3}^{+2}$\\
                & $\log \xi$ & -- & $2.77_{-0.08}^{+0.10}$\\
                & $A_{\rm Fe}$ & -- & $1.5_{-0.2}^{+0.4}$\\
                & $\log N$ & -- & [19] \\
                & \text{norm ($10^{-3}$)} & -- & $1.20_{-0.12}^{+0.08}$\\
\hline
\multicolumn{2}{r}{$\chi^2/\textrm{d.o.f.}$} & 1793/1690 = 1.1 & 1898/1687 = 1.1\\ \hline
\multicolumn{4}{c}{Cross normalisation factors} \\
&  $C_{\rm NICER}$ & [1.0] & [1.0]\\
& $C_{\rm \nustar-A}$ & $1.295_{-0.001}^{+0.009}$ & $1.301_{-0.007}^{+0.011}$\\
& $C_{\rm \nustar-B}$ & $ 1.264_{-0.001}^{+0.010}$ & $1.270_{-0.007}^{+0.010}$\\
&  $C_{\rm \ixpe-DU1}$ & $0.878\pm0.006$ & $0.861\pm0.006$\\
& $C_{\rm \ixpe-DU2}$ & $0.890\pm0.006$ & $0.870\pm0.006$\\
& $C_{\rm \ixpe-DU3}$ & $0.878\pm0.006$ & $0.860\pm0.006$\\ \hline
\multicolumn{4}{c}{Photon flux ratios in 2--8\,keV} \\
& $F_{\rm \texttt{diskbb}}/F_{\rm tot}$ & 0.63 & 0.66\\
& $F_{\rm \texttt{comptt}}/F_{\rm tot}$ &  0.35 & 0.31 \\
& $F_{\rm \texttt{gauss}, \texttt{relxillNS}}/F_{\rm tot}$ & 0.02 & 0.03\\ \hline
\end{tabular}
\tablefoot{The estimated unabsorbed flux in 2--8\,keV is ${\sim}1.2\times10^{-8}$\,\fluxcgs, corresponding to a luminosity of ${\sim}1.7\times10^{38}$\,\lum\ for a distance to the source of 11\,kpc \citep{Fender00}. The errors are reported at 90\%~CL.}
\end{table*}

\begin{figure} 
\centering
\includegraphics[width=0.95\linewidth]{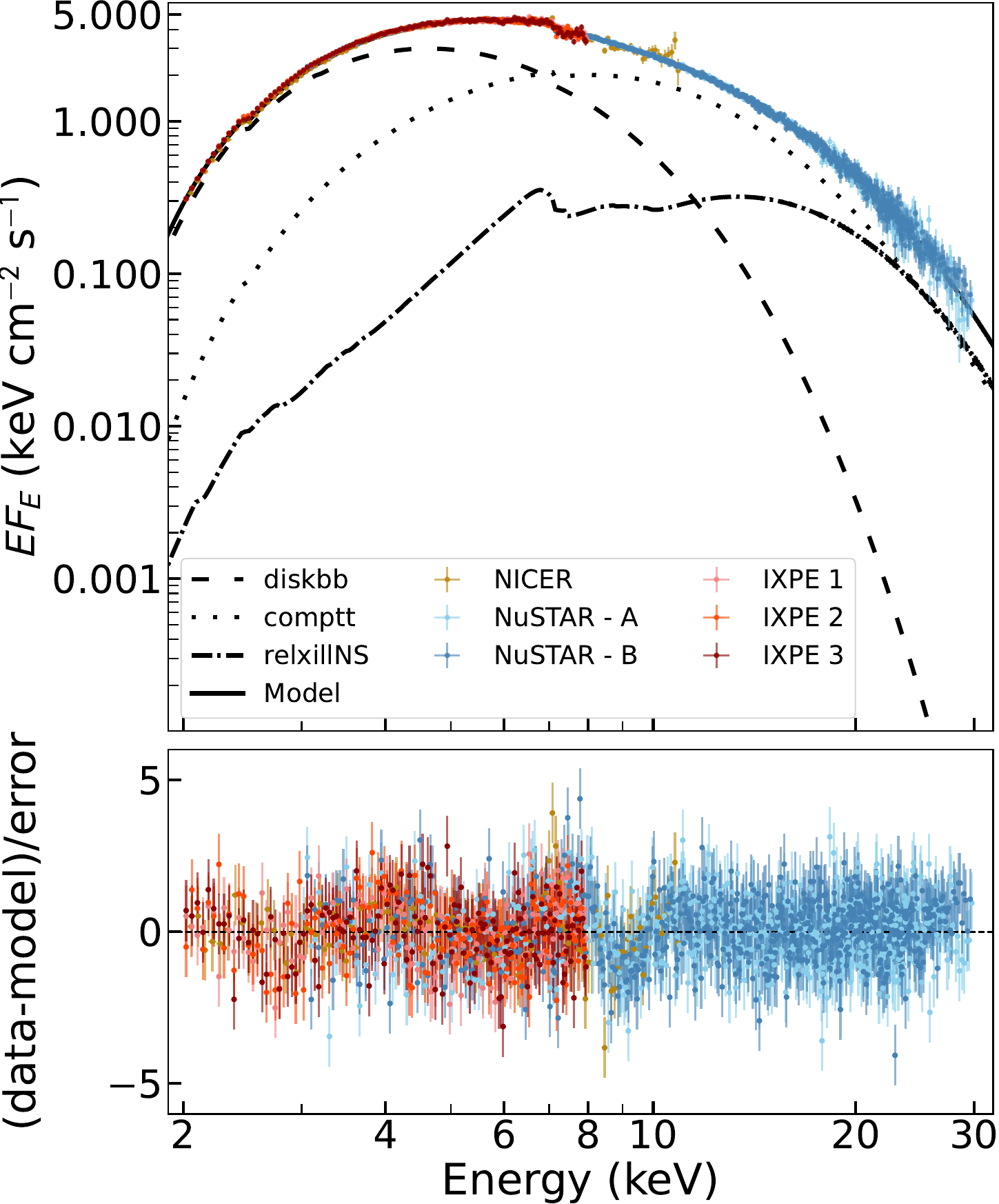}
\caption{Same as Fig.~\ref{fig:bestfitA} but for Model~B, with the dash-dotted lines showing the \texttt{relxillNS} component.} 
\label{fig:bestfitB}
\end{figure}

To better model the reflection component, we used the \texttt{relxillNS} model \citep{Garcia22} instead of the Gaussian; this model is hereafter referred to as Model~B: \texttt{tbabs*(diskbb+comptt+relxillNS)}. In this case, we froze the density parameter, $\log N$, at its maximum allowed value \citep{Garcia22}; the NS spin at 0 \citep{Galloway08, Miller_Miller16, Ludlam24}, the outer radius at $1000\,R_{\rm g}$; and the reflection fraction parameter at $-1$ to select only the reflection spectral component. We assumed the same temperature for the continuum blackbody in the \texttt{relxillNS} model and in the \texttt{comptt} seed photons; the assumption corresponds to an illumination originating from the BL and/or SL. The inner and outer emissivity indices were assumed to be equal, with the break radius not used. The best-fit values and errors were obtained from a Markov chain Monte Carlo (MCMC) with 100 walkers, a burn-in of $2\times10^5$, and a chain length of $4\times 10^5$, obtaining the values reported in Table~\ref{tab:spectrum}. The spectral energy distribution with residuals is shown in Fig.~\ref{fig:bestfitB}. The accretion column density is in line with the expected values for the interstellar medium absorption. The emissivity index is smaller than the one fixed in \cite{LaMonaca24gx} and \cite{Li25}, but the value is in agreement with the one obtained by \cite{Ludlam25}. The inner radius is slightly truncated but slightly smaller than the inferred one in HB by \cite{LaMonaca24gx}. 

Spectropolarimetric analysis for both Model~A and Model~B was performed using weighted spectra \citep{DiMarco_2022} in \textsc{xspec}. Firstly, we fitted only the \ixpe Stokes $I$ data with the model parameters frozen to the broadband best-fit parameters of Model~A, obtaining a $\chi^2/\textrm{d.o.f.} = 442/444 = 1.0$. Then, we performed spectropolarimetric analyses of the $I$, $Q$, and $U$ spectra to find a possible dependence of the polarisation with the energy. We started applying the \texttt{polconst} model, which assumes a constant polarisation with the energy, to Model~A, in the \ixpe energy band. The results are reported in Table~\ref{tab:spectropol_ModelA} and are compatible with the one obtained with the \texttt{pcube} model-independent analysis reported in Table~\ref{tab:PD_PAvsEnergy} in the 2--8\,keV energy band. Subsequently, we applied the \texttt{pollin} model, which assumes a polarisation that varies linearly with energy, and the \texttt{polpow} model, which assumes a polarisation that varies as a power law with energy. In both scenarios, we assumed a constant PA. The results of these different polarimetric models are reported in Table~\ref{tab:spectropol_ModelA}. The F-test between the \texttt{pollin} and \texttt{polconst} models gives F = 2.7, corresponding to a probability of improvement of 96\%, showing a preference for the \texttt{pollin} model compared to \texttt{polconst}. A similar indication is obtained by  Akaike's information criterion, which gives 772 for \texttt{polconst} and 767 for \texttt{pollin}, with a difference of 5 corresponding to a 'meaningful' improvement. The \texttt{polpow} model gives the same results of \texttt{pollin} with a negative $A_{\rm index}$, corresponding to increased PD with energy. This analysis confirms the presence of energy variability, even if we cannot disentangle between \texttt{pollin} and \texttt{polpow}, as is reported in Table~\ref{tab:PD_PAvsEnergy} and Figure~\ref{fig:PD_PAvsEnergy} from the model-independent \texttt{pcube} analysis.

\begin{table*}[!hbt]
\centering
\caption{Weighted spectropolarimetric analysis of \gx in the NB in the 2--8\,keV using Model~A combined with different polarisation models: \texttt{polconst}, \texttt{pollin}, and \texttt{polpow}.}
\label{tab:spectropol_ModelA}
\begin{tabular}{lccc}
\hline \hline
& Model~A * \texttt{polconst} &  Model~A * \texttt{pollin} & Model~A * \texttt{polpow}\\
\hline
 PD / $A_1$  (\%)$^{\dagger}$ & $1.2\pm0.3$  & $-0.057\pm0.008$ & $0.16_{-0.13}^{+0.43}$\\
 $A_{\rm slope}$ (\%\,keV$^{-1}$) & -- & $0.4\pm0.3$ & -- \\
$A_{\rm index}$  & -- & -- & $-1.4_{-1.0}^{+0.9}$ \\
PA / $\psi_1$  (deg)$^{\dagger}$ & $35\pm7$ & $36\pm6$ & $37\pm6$ \\
$\psi_{\rm slope}$ (deg\,keV$^{-1}$) & -- & [0] & -- \\
$\psi_{\rm index}$  & -- & -- & [0]\\
\hline
$\chi^2$/dof &  768/737=1.04 & 761/735=1.03 & 761/735=1.03\\ 
\hline
\end{tabular}
\tablefoot{The errors are reported at 90\%~CL.\\
$^{\dagger}$For the \texttt{pollin} and \texttt{polpow} models, $A_1$ and $\psi_1$ refer to 
the PD and PA values at 1\,keV.}
\end{table*}

\begin{figure}
   \centering
   \includegraphics[width=0.9\columnwidth]{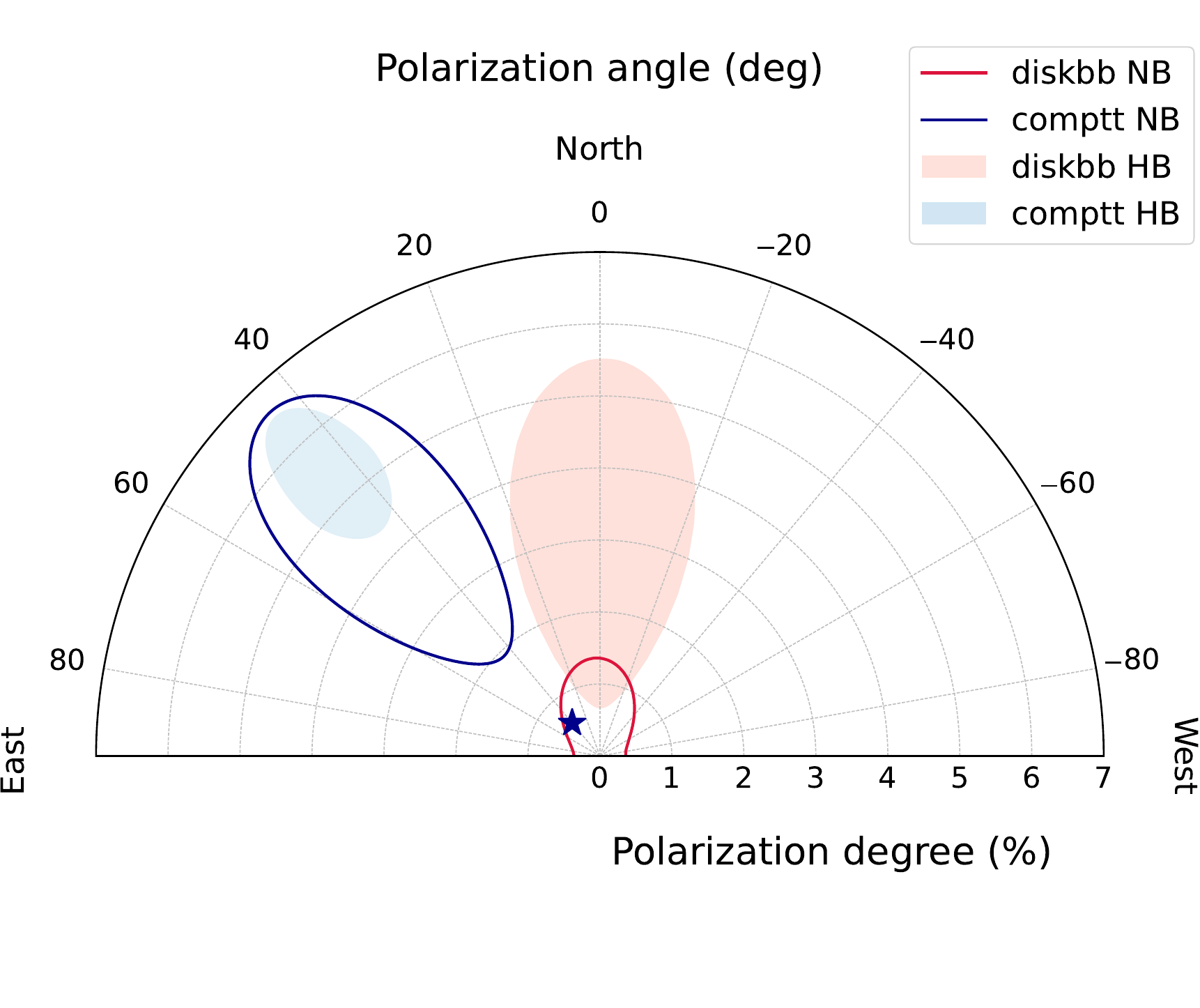}
   \caption{Polar plot with the results of the spectropolarimetric analysis for \gx in the NB and in the HB: a \texttt{polconst} component is associated with the \texttt{diskbb} and \texttt{comptt} components separately, without any prior constraints on the PA. The contours represent the allowed regions at 90\%~CL. The colour-shaded contours for the HB are from \cite{LaMonaca24gx}. The dark blue star represents the maximum polarisation expected by a SL \citep{Bobrikova25}; see Sect.~\ref{sec:discussion}. \label{fig:contours_spectropol}}
\end{figure}

Considering that soft and hard components are expected to have different polarisations, we also performed a weighted spectropolarimetric analysis associating the polarisation with each spectral component. It is worth noting that fluorescence emission processes, such as the Gaussian line, are expected to be almost unpolarised \citep[see, e.g.,][]{Churazov02, Ingram23, Veledina24}, while the reflection continuum is expected to be highly polarised \citep{Matt93,Poutanen96}. Thus, in Model~A, we prescribed polarisation to the disc and the Comptonised components, associating a constant polarisation, \texttt{polconst}, with each spectral component and keeping the Gaussian line unpolarised. This approach gives rise to $\chi^2/\textrm{d.o.f.}=759/734=1.03$ and the results reported in Fig.~\ref{fig:contours_spectropol} and Table~\ref{tab:spectropol_modelA_2polconts}. 

\begin{table}
\centering
\caption{Weighted spectropolarimetric analysis of \gx in the NB and HB using Model~A of Table~\ref{tab:spectrum} with a different constant polarisation associated with each spectral component.}
\label{tab:spectropol_modelA_2polconts}
\begin{tabular}{lrcc}
\hline \hline
Spectral& &  \gx NB &  \mbox{GX~340$+$0} HB$^{*}$\\
component& & &\\ 
 \hline
\texttt{diskbb} & PD (\%) &  <1.2  & $3.1 \pm 1.7$ \\ 
 & PA (\degr)  & -- &  $-1 \pm 16$\\ 
 \hline
\texttt{comptt}& PD (\%) &  $4.3 \pm 1.8$ & $5.2 \pm 1.0$\\ 
 & PA (\degr) & $44 \pm 13$ & $44 \pm 5$\\
\hline
\end{tabular}
\tablefoot{The errors are reported at 90\%~CL. A polarisation with zero amplitude is associated with the \texttt{Gaussian} component.\\
$^{*}$Spectropolarimetric results for \gx as obtained for the HB by \cite{LaMonaca24gx}.}
\end{table}

To assess the polarisation of the reflection, we also need to model its continuum, as was done in Model~B. Using this model, we associated a constant polarisation with each spectral component, obtaining a $\chi^2/\textrm{d.o.f.}=769/732=1.05$, and only an upper limit for the polarisation associated with each of the three spectral components at 90\%\,CL: $\textrm{PD}{<}1\%$ for the \texttt{diskbb}, $\textrm{PD}{<}8\%$ for \texttt{comptt}, and $\textrm{PD}{<}71\%$ for \texttt{relxillNS}. In order to constrain the polarisation of the reflection, we assumed that the polarisation of the disc is 0.7\%, as is expected when the inclination is $35\degr$ \citep{Chandrasekhar60, Sobolev63, Loktev22} and the polarisation of the Comptonised component is at the value obtained in Table~\ref{tab:spectropol_modelA_2polconts} for NB, obtaining an upper limit $\textrm{PD}<26\%$ for \texttt{relxillNS}. Fixing the Comptonised PD to ${<}0.6\%$ -- the maximum expected value, assuming the Comptonised emission from an optically thick and relatively narrow SL and using the parameters of the Comptonising
medium reported in Table~\ref{tab:spectrum}, namely $i=34\degr$ and $\rm kT=2.94\,keV$ \citep{Bobrikova25} -- we derived for the reflection component $\textrm{PD}=37_{-16}^{+18}\%$ and $\textrm{PA}=45_{-15}^{+11}$\,degr. 

\section{Discussion and conclusions} \label{sec:discussion}

\gx was observed by \ixpe twice. In the first observation, the source was in the HB \citep{LaMonaca24gx,Bhargava24}, and a PD of ${\sim}4\%$ was detected at high significance. In the second observation, reported here, the source moved between NB and HB, passing through the HA (see Figs.~\ref{fig:IXPE_HID} and \ref{fig:NICER_NuSTAR_HID}). This variability is confirmed by the \ixpe light curve and HR (see Fig.~\ref{fig:IXPE_LC_HR_PD_PA}a,b) and by the simultaneous EP-WXT and \nicer observations. The period in which \gx is in the HB corresponds roughly to the middle of the observation \citep[see also][where the same observation is studied]{Bhargava24b}. 

The results obtained from the model-independent polarimetric analysis performed in 2--8\,keV with the \texttt{ixpeobssim} package for the HB and the NB of \gx are reported in Fig.~\ref{fig:hb_nb}. Combining the new data during the HB with those already available, we obtain a marginal improvement in the significance of the detection of polarisation for the HB: $\textrm{PD} = 3.7\%\pm0.4\%$ with $\textrm{PA} = 36\degr\pm2\degr$ at 68\%~CL. When only NB is selected, we obtain the polarisation measurement for this state to be: $\textrm{PD} = 1.4\%\pm0.3\%$ with $\textrm{PA} = 37\degr\pm5\degr$ at 68\%~CL. The result is in agreement with the one reported by \cite{Bhargava24b}. Therefore, we observed a significant decrease in polarisation between the HB and the NB, as has been observed in other Z sources, and with PD values for each of the two branches in agreement with values measured in the same branch in other Z sources \citep[see, e.g.][for a review]{DiMarco25b}. In fact, \ixpe measured a polarisation of $1.8\% \pm 0.3\%$ with a PA of 140\degr$\pm$4\degr\ for \mbox{Cyg~X-2} in the NB \citep{Farinelli23},  PD=$1.8\%\pm0.4\%$ with a PA=$9\degr\pm2\degr$ in the NB+FB of \mbox{GX~5$-$1} \citep{Fabiani24}, and 
an upper limit of 1.5\% at 99\%~CL for the PD  
in the NB of \mbox{XTE~J1701$-$462}. Moreover, while the average PD decreases when the source changes state, the average PA remains consistent between NB and HB in all of them, including this new study for \gx. 

We also performed a time-resolved study of the polarisation, dividing the second observation into ten equal time bins of about 9.5~h each (see Fig.~\ref{fig:IXPE_LC_HR_PD_PA}c,d). This study shows a marginal indication of time variability at the beginning of the observation, consistent with the transition from HB to NB, but no evidence of variability during the flux drop. This is probably due to a lack of significance in the analysis and to the mixing of data from NB and HB when dividing the observation into equally spaced time bins, while the selection in the HID (see Fig.~\ref{fig:IXPE_HID}) allows for a better separation of the HB and the NB data, which show different polarisation. Moreover, no variation in the PA is observed during the NB, in contrast to the case of \mbox{XTE~J1701$-$462} that showed variations of the PA at 90\%~CL in the  NB \citep{DiMarco25a, Zhao25}. An energy-resolved polarisation study reported in Fig.~\ref{fig:PD_PAvsEnergy} shows an overall trend for PD similar to that observed for HB \citep{LaMonaca24gx,Bhargava24}, even if the maximum PD reached with energy is lower: $\textrm{PD}=4.0\% \pm1.5\%$ in the 7.0--8.0\,keV of the NB, compared to $\textrm{PD}=12\% \pm3\%$ in the 7.5--8.0\,keV of the HB \citep{LaMonaca24gx}. On the other hand, the PA in the NB does not show any rotation with the energy, unlike the HB, in which a variation of ${\sim}40\degr$ is observed between the soft and high energy bins \citep{LaMonaca24gx}. A similar trend with energy was also reported for the Sco-like source \mbox{GX~349$+$2} \citep{LaMonaca25gx349}. Moreover, an increase in the PD with energy was observed for some atoll sources, such as \mbox{4U 1820$-$303} \citep{DiMarco23b} and \mbox{GX~9$+$1} \citep{Prakash25}.

Combining the simultaneous observations by \ixpe and \nicer with a non-simultaneous observation by \nustar in the NB (see Table~\ref{tab:exposure} and Fig.~\ref{fig:NICER_NuSTAR_HID}), we performed a broadband spectral analysis, following the same approach as in \citet{LaMonaca24gx}. We used a simplified model consisting of a \texttt{diskbb} component for the soft emission, \texttt{comptt} for the hard Comptonised emission, and a Gaussian line to describe the main feature of the reflection component. This model provided an acceptable reduced $\chi^2$ (see Model~A in Table~\ref{tab:spectrum} and Fig.~\ref{fig:bestfitA}). The fit shows a spectrum dominated by the soft disc component with a relative flux of ${\sim}63\%$ of the total flux, while the Comptonised component has a relative flux of ${\sim}35\%$, and the Gaussian line is $\sim$2\% of the total. This is in contrast to the case of HB, where the spectrum was dominated by the Comptonised component (${\sim}78\%$ of the total flux), while the soft component only contributed ${\sim}21\%$ \citep{LaMonaca24gx}. Firstly, we performed a spectropolarimetric analysis, fixing the best-fit model for the Stokes $I$ spectrum to the values reported for Model~A in Table~\ref{tab:spectrum}, and applying a constant polarisation to it in the 2--8\,keV energy band, obtaining results consistent with those derived by model-independent analysis in the same energy band; see Table~\ref{tab:PD_PAvsEnergy}~and~\ref{tab:spectropol_ModelA}. We also tested polarisation models that are not constant with energy, finding that the \texttt{pollin} and \texttt{polpow} models are equally favoured with respect to \texttt{polconst} with a probability of 96\%, which confirms a dependence of polarisation on energy. Both polarimetric models report that PD grows in energy, while PA is constant with energy. No evidence of a PA 90\degr\ swap, corresponding to a change in the PD sign, is observed either. This energy dependence was also shown by model-independent analysis and has already been observed in the HB. In fact, applying the \texttt{pollin} model to the HB data reported in \citet{LaMonaca24gx} we obtained: $A_{1}=(1.9\pm1.3)$\%, $A_{\rm slope}=(0.6\pm0.4)$\%\,keV$^{-1}$, $\psi_{1}=21\degr\pm10\degr$, and $\psi_{\rm slope}=(4\pm3)$\degr\,keV$^{-1}$ with $\chi^2/\textrm{d.o.f.} = 1230/1335=0.9$. Although the polarisation in HB and NB has a similar slope and the PA in HB seems to have a not-zero slope, the higher average polarisation is due to the fact that at A$_1$ is higher in HB than in NB, confirming the indication in Figure~\ref{fig:contours_spectropol} of a soft component with higher polarisation in HB than in NB; further observations with a higher significance could be useful in the future to confirm these results.

Consequently, we performed a spectropolarimetric analysis to disentangle the polarisation properties of each spectral component. We assumed the Gaussian to be unpolarised (see Sect.~\ref{sec:spec}) and associated a \texttt{polconst} model with both the soft and the Comptonised components, obtaining the results reported in Table~\ref{tab:spectropol_modelA_2polconts}. Fig.~\ref{fig:contours_spectropol} shows this result in the polar plane of PD and PA, in which we also report the results for the spectropolarimetric analysis of HB as in \cite{LaMonaca24gx}: $\textrm{PD} =  3.1\% \pm 1.7\%$ with a $\textrm{PA} = -1\degr \pm 16\degr$ for the \texttt{diskbb} and $\textrm{PD} = 5.2\% \pm 1.0\%$ with a $\textrm{PA}=44\degr \pm 5\degr$ for \texttt{comptt}. Therefore, during changes in the source state, that is, when passing from HB to NB, the PD of the Comptonised component remains compatible at a confidence level higher than 90\%~CL without any significant variation in the PA, while the PD of the soft component (even if marginally in agreement with the value in the HB) drops below 1.2\% and the PA becomes unconstrained at 90\%~CL. This means that the average PD decreases in the NB with respect to the HB because the flux in the soft component, which is less polarised, is higher than in the HB. Similar results were reported for the Comptonised component of \mbox{GX~5$-$1} in which a PD at a level of ${\sim}4-5\%$ was obtained in the HB and in the NB+FB with compatible values for the PA \citep{Fabiani24}. 

In addition, we adopted a complete reflection model, \texttt{relxillNS}, to properly describe this spectral component, as is reported for the HB in \cite{LaMonaca24gx}. We obtained the best-fit parameters reported as Model~B in Table~\ref{tab:spectrum} and Fig.~\ref{fig:bestfitB}. As has already been reported for the HB \citep{LaMonaca24gx}, also in the case of the NB, it is not possible to disentangle the polarisation of the different spectral components of Model B. To determine the polarisation for the reflection component, we must make assumptions about the polarisation of the soft and hard spectral components. Assuming for the disc component a polarisation at 0.7\%, as is expected by \cite{Chandrasekhar60}, for the inclination of ${\sim}35$\degr, with null polarisation for the Comptonised component, gives rise to polarisation for the reflection component, which reaches unphysical values of up to 80\% \citep{Matt93,Poutanen96}. If we increase the PD of the Comptonised component, the reflection component shows a much lower polarisation, and to obtain values of ${\sim}20-30\%$, the Comptonised component needs to have PD${>}1$\%.

Even if spectropolarimetric analysis shows that the PA of the disc component is not constrained at a 90\%~CL, the marginal indication is that its PA is aligned with the one previously measured in HB, which is not orthogonal to that of the Comptonised component. This result is in contrast with theoretical expectations \citep[see, e.g.,][]{st85, Tomaru2024} for an optically thick Comptonising medium and it is possibly due to a misalignment between the disc and BL axis and the NS axis \citep{Bobrikova24a, Rankin24, LaMonaca24gx}. The PD of the disc component is expected to be ${<}0.7$\% when the inclination is 35\degr\ \citep{Chandrasekhar60, Sobolev63, Loktev22}, which is compatible with our results. In fact, the reflection model allows us to obtain an inclination of ${\sim}34\degr$ that is compatible with recent results reported in \citet{LaMonaca24gx} and \citet{Li25}, but also with previous results reported in the literature \citep{Dai09, Cackett10, Miller16}.

The PD of the Comptonised component remains constant at $\textrm{PD}{\sim}4\%$ between the HB and the NB. If we assume that the Comptonised emission comes from an optically thick and relatively narrow SL, we can get an upper limit on the PD of the harder component. Taking into account the parameters of the Comptonising medium of Tab.~\ref{tab:spectrum}, namely $i=34\degr$ and $\rm kT=2.94\,keV$, we expect a PD of ${<}0.6$\% \cite{Bobrikova25} (see Fig.~\ref{fig:contours_spectropol}). X-rays from such an SL are expected to be polarised along the NS rotation axis. If the SL is wider than we assume, it could only decrease PD, and the PA would remain unchanged \citep {Bobrikova25}. Other theoretical studies also provide a relatively low expected PD: ${<}2$\% for an optically thick BL and/or SL \citep{st85,Farinelli24}. We conclude that the BL and/or SL alone cannot explain the observed PD, which is significantly higher than predicted. One possible explanation for this high polarisation could be the scattering in the wind \citep{nitindala25}. We note that such an explanation agrees with the results for \gx from \citet{Miller16} in which the presence of wind is claimed, although this feature is not observed in the recent XRISM observation \citep{Ludlam25}. An alternative explanation for the observed higher polarisation is the presence of an ADC \citep{Church06}, which may contribute as an additional component of polarised emission \citep{DiMarco25b}. The ADC scenario also agrees with recent results reported in \cite{Ludlam25}.

This new observation and the previous ones from other WMNSs are helping us to better characterise the polarimetric properties of the accretion flow and the geometry of the emission regions in these systems. A general picture is emerging: the hard state (HB) appears to be highly polarised \citep{Cocchi23, Fabiani24, LaMonaca24gx, DiMarco25a} and probably dominated by emission from an extended corona or wind, whereas the softer states (NB and FB) are probably dominated by disc and BL emissions \citep{Rankin24}. In the future, the development of new models and the improved statistics achievable by observing the same sources again for a longer time will help to further unveil the points that at present remain unanswered.

\begin{acknowledgements}
This work is supported by National Key R\&D Program of China (grant No. 2023YFE0117200), and National Natural Science Foundation of China (grant No. 12373041 and No. 12422306), and Bagui Scholars Program (XF).
This research used data products provided by the IXPE Team (MSFC, SSDC, INAF, and INFN) and distributed with additional software tools by the High-Energy Astrophysics Science Archive Research Center (HEASARC), at NASA Goddard Space Flight Center (GSFC). The Imaging X-ray Polarimetry Explorer (IXPE) is a joint US and Italian mission.  
This work is based on data obtained with the Einstein Probe, a space mission supported by the Strategic Priority Program on Space Science of the Chinese Academy of Sciences, in collaboration with ESA, MPE and CNES (Grant No. XDA15310000, No. XDA15052100).
The authors acknowledge the NICER and NuSTAR team whose data were used in this research.
The Italian contribution is supported by the Italian Space Agency (Agenzia Spaziale Italiana, ASI) through contract ASI-OHBI-2022-13-I.0, agreements ASI-INAF-2022-19-HH.0 and ASI-INFN-2017.13-H0, and its Space Science Data Center (SSDC) with agreements ASI-INAF-2022-14-HH.0 and ASI-INFN 2021-43-HH.0, and by the Istituto Nazionale di Astrofisica (INAF) and the Istituto Nazionale di Fisica Nucleare (INFN) in Italy. FLM and ADM are partially supported by MAECI with grant CN24GR08 “GRBAXP: Guangxi-Rome Bilateral Agreement for X-ray Polarimetry in Astrophysics”. 
AB is supported by the Finnish Cultural Foundation grant No. 00240328.
RML and SL acknowledge support by NASA under grant No. 80NSSC23K0498.

\end{acknowledgements}

%
%
\bibliographystyle{yahapj}
\bibliography{biblio}
-------------------------------------------------------------

\label{LastPage} 
\end{document}